\documentclass[preprint2,10.5pt]{aastex}
\usepackage{epsfig}
\begin{document}
\title{\large{\rm{HDE 344787, THE POLARIS ANALOGUE THAT IS EVEN \\ MORE INTERESTING THAN POLARIS}}}
\author{David G. Turner$^{1,2,6,7,8}$, Daniel J. Majaess$^{1,2}$, David J. Lane$^{1,2}$, John R. Percy$^{3,6}$ \\ Darlene A. English$^{4,7}$, Richard Huziak$^5$}
\affil{$^1$ Saint Mary's University, Halifax, Nova Scotia, Canada.}
\affil{$^2$ The Abbey Ridge Observatory, Stillwater Lake, Nova Scotia, Canada.}
\affil{$^3$ Erindale College, University of Toronto, Erindale, Ontario, Canada.}
\affil{$^4$ Sir Wilfred Grenfell College, Memorial University, Corner Brook, Newfoundland, Canada.}
\affil{$^5$ SED Systems, Saskatoon, Saskatchewan, Canada.}
\affil{$^6$ Visiting Astronomer, Kitt Peak National Observatory, National Optical Astronomy Observatories.}
\affil{$^7$ Visiting Astronomer, Dominion Astrophysical Observatory, Herzberg Institute of Astrophysics, \\ National Research Council of Canada.}
\affil{$^8$ Visiting Astronomer, Harvard College Observatory Photographic Plate Stacks.}
\email{\rm{turner@ap.smu.ca}}

\begin{abstract}
A collection of active photometric observations over the last half decade, archival data from the past 120 years, radial velocity observations from 1984, and recent monitoring through a pro-am collaboration reveal that the 9th magnitude F9 Ib supergiant HDE 344787 is a double-mode Cepheid variable of extremely small amplitude. It displays remarkably similar, but much more extreme, properties to the exotic Cepheid Polaris, including a rapidly-increasing period and sinusoidal light variations of decreasing amplitude suggesting that pulsational stability may occur as early as 2045. Unlike Polaris, HDE 344787 displays sinusoidal light variations at periods of both $5^{\rm d}.4$ and $3^{\rm d}.8$ days, corresponding to canonical fundamental mode and overtone pulsation. But it may be similar to Polaris in helping to define a small subgroup of Cepheids that display characteristics consistent with a first crossing of the instability strip. An update of 2010 observations of this remarkable star is presented.
\end{abstract}

\section{Introduction}

\begin{figure}[h]
\centerline{
\epsfig{file=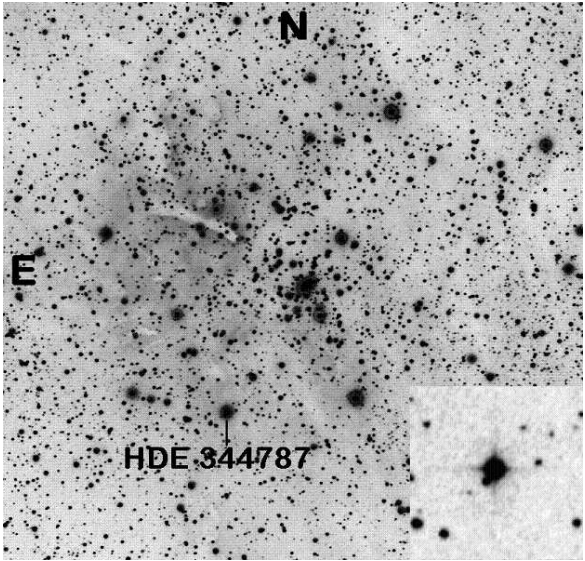, scale=0.35}
}
\caption{\small{The $32^{\prime} \times 32^{\prime}$ field of the young cluster NGC 6823 and its H II region, showing the location of HDE 344787. The inset is an enlargement showing the close ($11^{\prime\prime}$) B-type companion to HDE 344787.}}
\label{fig1}
\end{figure}

  HDE 344787 (BD$+22^{\circ}3786$) is unusual in being an F9 Ib-II supergiant (Turner 1979; Shi \& Hu 1999) lying on the outskirts (Fig.~\ref{fig1}) of the $2\times10^6$ year-old cluster NGC 6823 (Guetter 1992), but with a high membership probability from proper motions (Erickson 1971). An evolved intermediate-mass star in such close proximity to a very young cluster must be a spatial coincidence, which is testable by radial velocities. It has an optical companion $11^{\prime\prime}$ southeast and six magnitudes fainter that is of probable B spectral type (Massey et al. 1995).

  All-sky and differential photoelectric observations for the star were obtained in 1978, 1979, 1980, and 1981 with the 0.4m telescope at Kitt Peak National Observatory (KPNO) in a search for potential Cepheid-like variability, and radial velocity measures were obtained at the Dominion Astrophysical Observatory (DAO) in 1984. Initial examination of the data (Turner 1979) suggested a 4-day variation in brightness, although the observations are also consistent with light constancy! Tycho observations of the star (ESA 1997) are of low quality and were averaged. Additional data were obtained from the ASAS-3 survey (Pojma{\'n}ski 2002), from recent CCD imaging, mainly from the Abbey Ridge Observatory (ARO), and from examination of images in the Harvard College Observatory Photographic Plate Collection (HCO). Small brightness changes on the HCO plates were detected at the $0^{\rm m}.05$ level, tied to a surrounding tight sequence of three stars of almost identical brightness and colour to HDE 344787, one slightly brighter and the other slightly fainter, thereby eliminating problems arising from differential extinction during plate exposures, which typically limits the accuracy attainable from eye estimates off photographic plates to $\pm0^{\rm m}.1$ or worse.

  Absolute photometry of HDE 344787 from CCD imaging is rather challenging because of the star's small light amplitude. Since the reference and check stars for the Cepheid differ in colour from HDE 344787, it was necessary to account for atmospheric effects on the derived magnitude estimates. The procedure for that outlined by Turner et al. (2009) was followed.

\section{Observational Results}

\begin{figure}[h]
\centerline{
\epsfig{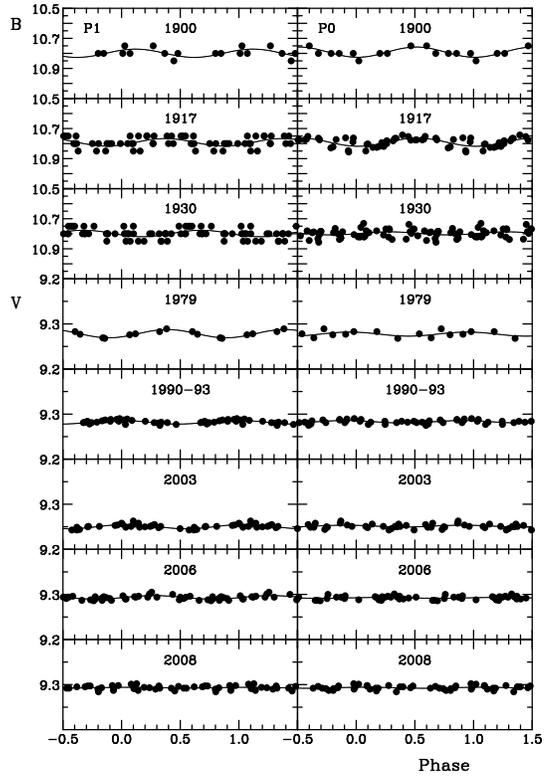}
}
\caption{\small{Sample light curves of HDE 344787 for FM (F0) and OT (F1) pulsation, derived from HCO estimates (1900, 1917, 1930), from KPNO observations (1979), from Tycho magnitudes (1990-93), from ASAS-3 (2003), and from the ARO (2006, 2008). Note the changing light amplitude over the past century.}}
\label{fig2}
\end{figure}

\begin{figure}[h]
\centerline{
\epsfig{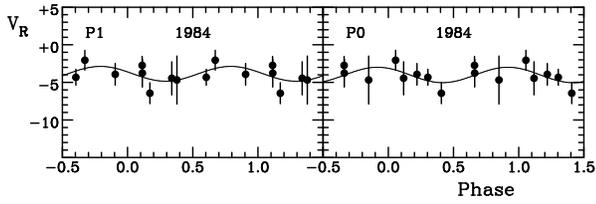}
}
\caption{\small{Radial velocity curves of HDE 344787 from observations with the DAO 1.85m telescope (1984).}}
\label{fig3}
\end{figure}

  The results were unexpected. Sample light curves for the Cepheid are shown in Fig.~\ref{fig2} along with fitted sine waves, indicating the low level of light variability at both photographic ({\it B}) and visual ({\it V}) wavelengths. The ``variability'' is presently only marginally larger than the scatter observed in reference stars of constant brightness (e.g., ASAS-3 2003, ARO 2006, 2008), was larger 30 years ago (e.g., KPNO 1979), and a century earlier was large enough to detect on photographic plates (e.g., HCO 1900, 1917, 1930). The variability is also seen in closely-adjacent Tycho observations combined into weighted means (e.g., 1990-93), and in radial velocity measures (DAO 1984, Fig.~\ref{fig3}).

\begin{figure}[h]
\centerline{
\epsfig{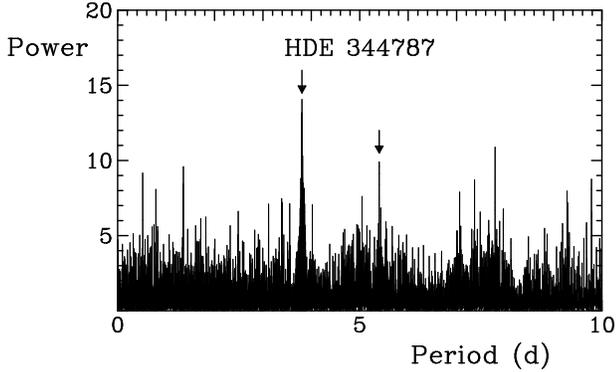}
}
\caption{\small{Fourier analysis of the power spectrum of HDE 344787 from HCO estimates, with peaks for OT pulsation (left arrow) and FM pulsation (right arrow).}}
\label{fig4}
\end{figure}

  Fourier analysis of the magnitude estimates from the 1890-1951 Harvard plates, shown in Fig.~\ref{fig4}, produced a dominant signal at $P=3^{\rm d}.8$, representing first overtone (OT) pulsation in the Cepheid, and a secondary peak at $P=5^{\rm d}.4$ (with aliases omitted), corresponding to fundamental mode (FM) pulsation. The $5^{\rm d}.4$ period went unnoticed in the photoelectric observations because of limited durations for the observing runs.

\begin{figure}[h]
\centerline{
\epsfig{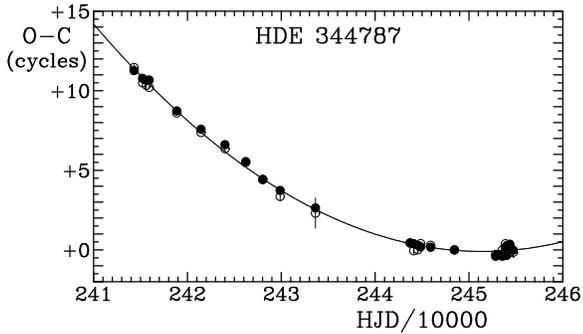}
}
\caption{\small{O--C variations of HDE 344787 for FM (open circles) and OT (filled circles) pulsation.}}
\label{fig5}
\end{figure}

  A working ephemeris was constructed for HDE 344787 using maxima obtained from phasing the Tycho data as zero-points:
\begin{displaymath}
\mathrm{HJD_{max}=2448449.4642+5.4019 {\it E}\,\,(for\,\,P0)}
\end{displaymath}
\begin{displaymath}
\mathrm{HJD_{max}=2448447.4505+3.8011 {\it E}\,\,(for\,\,P1)}
\end{displaymath}
the stronger signal,
where {\it E} is the number of elapsed cycles. The ephemerides were used to phase the observations in an O--C analysis presented in Fig.~\ref{fig5}, although the individual light curves were phased using periods appropriate for the epochs of observation, given that the period is changing so rapidly. The HCO data from the early 1900s indicate that the phase of light maximum progresses through a complete cycle in less than a decade, with both periods increasing in phase at the same rate.

\begin{figure}[h]
\centerline{
\epsfig{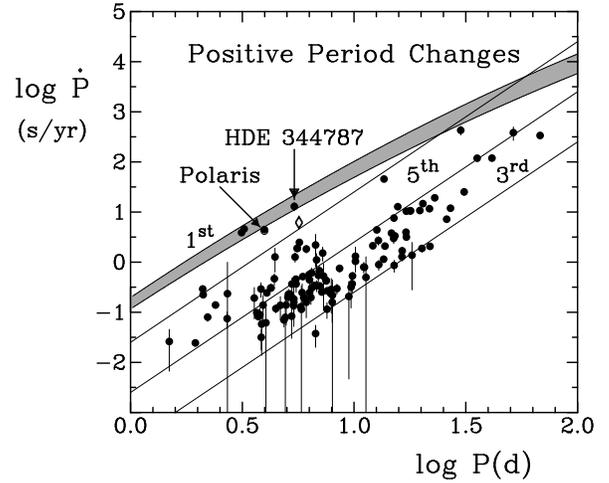}
}
\caption{\small{Rate of period change as a function of period for Cepheids with established period increases. Note the agreement of the observed rates for putative first crossers with expectations (gray region).}}
\label{fig6}
\end{figure}

A regression analysis of the combined data sets is affected only weakly by the scatter in the O--C data for recent epochs, a result of the rapidly weakening light amplitude. A best fit parabolic trend implies a period increase for HDE 344787 of $12.96 \pm2.41$ s yr$^{-1}$ for P0. The rapid rate of period increase in HDE 344787 corresponds exactly with expectations from previous studies of Cepheid period changes (Turner et al. 2006), and predictions from stellar evolutionary models, for a star of solar metallicity crossing the instability strip for the first time (Fig.~\ref{fig6}, see also Turner 2009). Temporal variations in light amplitude $\Delta V$ (with $\Delta B$ assumed equal to $1.5\times \Delta V$ for comparison purposes) reveal a steady decrease in light amplitude (Fig.~\ref{fig8}) that may represent the signature of a star about to leave the Cepheid instability strip.

\begin{figure}[h]
\centerline{
\epsfig{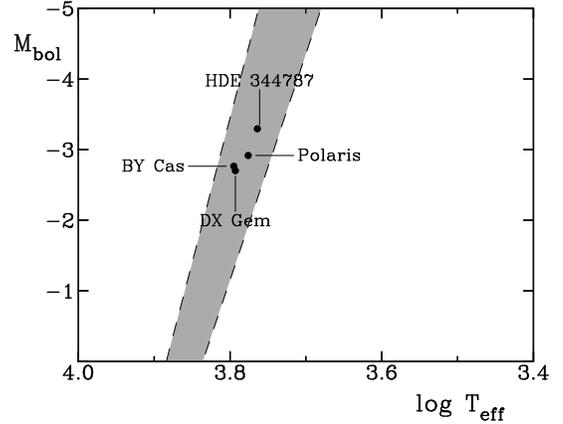}
}
\caption{\small{The location of putative first crossing Cepheids in the instability strip (gray region).}}
\label{fig7}
\end{figure}

\begin{figure}[h]
\centerline{
\epsfig{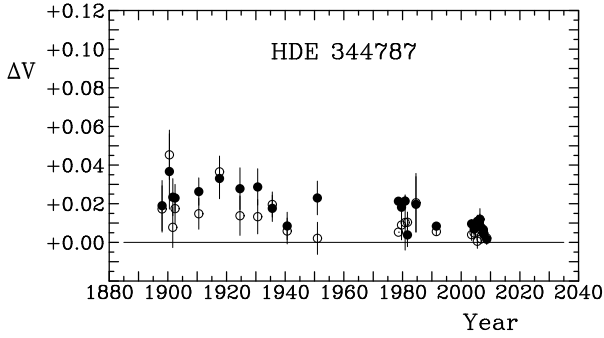}
}
\caption{\small{Light amplitude for HDE 344787 as a function of time for FM (open circles) and OT (filled circles) pulsation.}}
\label{fig8}
\end{figure}

\begin{figure}[h]
\centerline{
\epsfig{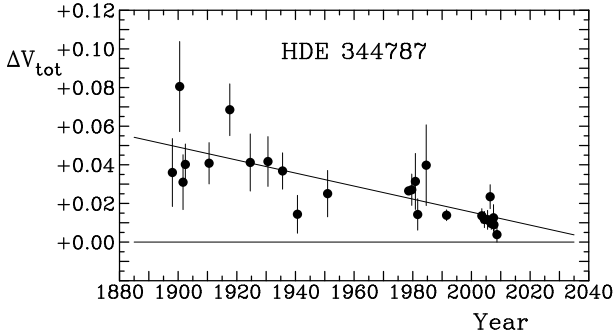}
}
\caption{\small{Total light amplitude for pulsation in HDE 344787 (FM~+~OT) as a function of time. The linear relation depicts the long-term declining trend.}}
\label{fig9}
\end{figure}

  The strongest signal seems to change between modes temporally, with the overtone mode dominating, but the total pulsation energy, as represented by a combination of the luminous signals into $\Delta V_{\rm tot}$ (Fig.~\ref{fig9}) is clearly waning. The star may cease to pulsate entirely by 2045, although the signal was barely detectable during the 2008-2009 observing seasons, and may have already been damped out by convection.

\section{Comparison With Polaris}

\begin{figure}[h]
\centerline{
\epsfig{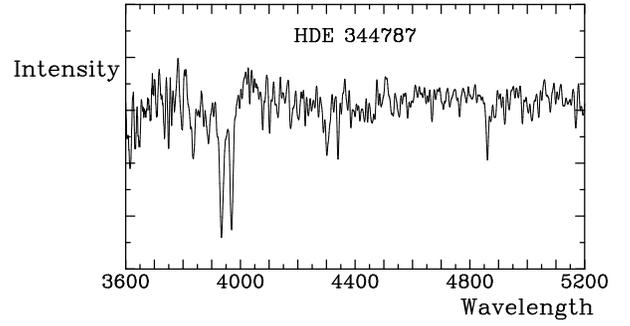}
}
\caption{\small{A spectrum of HDE 344787 taken at the DAO at a dispersion of 60 \AA\,mm$^{-1}$.}}
\label{fig10}
\end{figure}

  The following points can be noted when comparing HDE 344787 with Polaris, another F-type supergiant with a period of $\sim 4$ days that has a small light amplitude and may be crossing the instability strip for the first time: (i) HDE 344787 has a smaller light amplitude than Polaris, making it more difficult to observe, (ii) its rate of period increase of $12.96 \pm2.41$ s yr$^{-1}$ is $\sim 3$ times faster than that of Polaris ($4.5$ s yr$^{-1}$, but consistent with its longer fundamental mode pulsation period (P0 = 5$^{\rm d}$.4 versus 4$^{\rm d}$), (iii) HDE 344787 is a double-mode pulsator, unlike Polaris, for which no secondary period is evident, (iv) the intrinsic colour for a F9 Ib-II supergiant (HDE 344787, Fig.~\ref{fig10}) is {\it (B--V)}$_0$ = +0.63 (Kron 1978), versus {\it (B--V)}$_0$ = +0.58 for Polaris (Turner 2006), consistent with the difference in spectral type (Polaris is F7 Ib) and difference in pulsation period, (v) the light amplitude of HDE 344787 has been in steady decline since 1890, whereas Polaris had a steady but slow decline in light amplitude prior to its unusual ``glitch'' in 1963--66 (Turner et al. 2005; Turner 2009), and is presently recovering erratically from a minimum in 1988, (vi) the putative first-crossing Cepheids, HDE 344787, Polaris, DX Gem (P = 3$^{\rm d}$.1), and BY Cas (P = 3$^{\rm d}$.2) have intrinsic colours that place them near the centre of the instability strip, consistent with model expectations (Alibert et al. 1999).

\section{Discussion}

\begin{figure}[h]
\centerline{
\epsfig{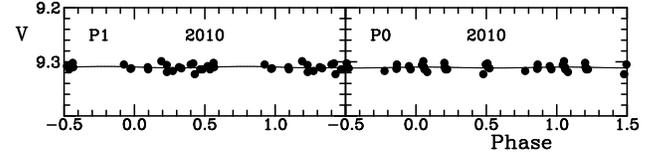}
}
\caption{\small{Recent ARO observations of HDE 344787, phased to FM (P0) and OT (P1) pulsation.}}
\label{fig11}
\end{figure}

Recent ARO observations of HDE 344787 (Fig.~\ref{fig11}) reveal a level of variability in 2010 not significantly different from zero: $\Delta V =0.002$ for FM pulsation and $\Delta V =0.003$ for OT pulsation, the normal dominant mode. At that level the star would be an ideal target for Canada's MOST space telescope, were it not for the fact that HDE 344787 lies outside the zone of continuous observation for MOST. As a consequence, it is difficult to determine if the star is still pulsating. Several recent sets of observations of the star have been made with facilities like the ARO and the campus telescope in Saskatoon (Huziak), so it is a project for which even keen amateur astronomers can contribute. The star is better placed for observation than Polaris (North Celestial Pole), but its extremely low light amplitude makes it a challenging object to study, as it has been for more than a century now.

\end{document}